\documentclass[preprint,aps,showpacs]{revtex4}

\usepackage{epsfig}

\begin{document}

\title{Dielectron production in $^{12}$C+$^{12}$C collisions at 2 AGeV with HADES}

\author{G.~Agakichiev$^{8}$, C.~Agodi$^{2}$, H.~Alvarez-Pol$^{16}$,
A.~Ba\l anda$^{3,d}$, D.~Bertini$^{4}$, J.~Bielcik$^{4}$ ,
G.~Bellia$^{2,a}$, M.~B\"{o}hmer$^{12}$, H.~Bokemeyer$^{4}$,
J.L.~Boyard$^{14}$
% ...
}

\affiliation{
(HADES Collaboration) \\
\mbox{$^{1}$ Institute of Physics, Slovak Academy of Sciences, 84228 Bratislava, Slovakia}\\
\mbox{ $^{2}$ Istituto Nazionale di Fisica Nucleare - Laboratori
Nazionali del Sud, 95125 Catania, Italy }\\
\mbox{$^{3}$ Smoluchowski Institute of Physics, Jagiellonian
University of Cracow, 30059 Cracow, Poland}\\
\mbox{$^{4}$ Gesellschaft f\"{u}r Schwerionenforschung mbH, 64291
Darmstadt, Germany}\\
\mbox{$^{5}$ Institut f\"{u}r Strahlenphysik,
Forschungszentrum Rossendorf, 01314 Dresden, Germany}\\
\mbox{$^{6}$ Joint Institute of Nuclear Research, 141980 Dubna, Russia}\\
\mbox{$^{7}$ Institut f\"{u}r Kernphysik, Johann Wolfgang
Goethe-Universit\"{a}t, 60486 Frankfurt, Germany}\\
\mbox{$^{8}$ II.Physikalisches Institut, Justus Liebig
Universit\"{a}t Giessen, 35392 Giessen, Germany}\\
\mbox{$^{9}$ Istituto Nazionale di Fisica Nucleare, Sezione di Milano,
20133 Milano, Italy}\\
\mbox{$^{10}$ Institute for Nuclear Research, Russian Academy of
Science, 117312 Moscow, Russia}\\
\mbox{$^{11}$ Institute of Theoretical and Experimental Physics,
117218 Moscow, Russia}\\
\mbox{$^{12}$ Physik Department E12, Technische Universit\"{a}t
M\"{u}nchen, 85748 Garching, Germany}\\
\mbox{$^{13}$ Department of Physics, University of Cyprus, 1678
Nicosia, Cyprus}\\
\mbox{ $^{14}$ Institut de Physique Nucl\'{e}aire d'Orsay,
CNRS/IN2P3, 91406 Orsay Cedex, France}\\
\mbox{$^{15}$ Nuclear Physics Institute, Academy of Sciences of
Czech Republic, 25068 Rez, Czech Republic}\\
\mbox{$^{16}$ Departamento de F\'{\i}sica de Part\'{\i}culas,
University of Santiago de Compostela,}\\
\mbox{15782 Santiago de Compostela, Spain}\\
\mbox{$^{17}$ Instituto de F\'{\i}sica Corpuscular, Universidad de
Valencia-CSIC, 46971 Valencia, Spain} \\
\\
\mbox{ $^{a}$ also at Dipartimento di Fisica e Astronomia,
Universit\`{a} di Catania, 95125, Catania, Italy}\\
\mbox{$^{b}$ also at Technische Universit\"{a}t Dresden, 01062
Dresden, Germany}\\
\mbox{$^{c}$ also at Dipartimento di Fisica, Universit\`{a} di
Milano, 20133 Milano, Italy}\\
\mbox{$^{d}$ also at Pa\'{n}stwowa Wy\.{z}sza Szko\l a Zawodowa,
33-300 Nowy S\c{a}cz, Poland}\\
\mbox{$^{e}$ also at Technische Universit\"{a}t Darmstadt, 64289
Darmstadt, Germany}\\
 \dag deceased
}

\begin{abstract}

The invariant-mass spectrum of $e^+ e^-$ pairs produced in
$^{12}$C+$^{12}$C collisions at an incident energy of 2 GeV per
nucleon has been measured for the first time. The measured pair
production probabilities span over five orders of magnitude from the
$\pi^0$-Dalitz to the $\rho/\omega$ invariant-mass region. Dalitz
decays of $\pi^0$ and $\eta$ account for all the yield up to 0.15
GeV/c$^2$, but for only about 50\% above this mass.  A comparison
with model calculations shows that the excess pair yield is likely
due to baryon-resonance and vector-meson decays. Transport
calculations based on vacuum spectral functions fail, however, to
describe the entire mass region.
\end{abstract}


\maketitle

The properties of hot and dense hadronic matter represent a key
problem in heavy-ion physics, with far-reaching implications for
other fields such as the physics of compact stars.  They are
governed by non-perturbative QCD and cannot be derived directly from
the underlying Lagrangian.  Models predict, however, that hadron
properties, such as mass and lifetime, depend on the temperature and
density of the medium.  While some hadronic many-body calculations
give a broadening of the meson in-medium spectral function, other
approaches predict dropping meson masses as precursors of chiral
symmetry restoration \cite{Rapp:1999ej}.

Experimentally, in-medium properties are difficult to observe.
Suitable probes are dileptons ($\mu^+\mu^-$ or $e^+e^-$) from decays
of short-lived resonances produced inside the hadronic matter
created in the course of relativistic or ultra-relativistic
heavy-ion collisions. At the CERN SPS, the CERES collaboration has
established a significant excess of the dielectron yield as compared
to that expected from the decays of hadrons after chemical
freeze-out in 40 and 158~AGeV Pb+Au collisions
~\cite{Agakichiev:2005ai}.  At 1~AGeV, the DLS collaboration found
unexpectedly large electron-pair yields in C+C and Ca+Ca
collisions~\cite{dls}.  In contrast to the situation at SPS beam
energies, the DLS results cannot be described satisfactorily within
the various scenarios proposed for possible changes of the in-medium
spectral functions~\cite{Rapp:1999ej,weise,rhospectral}.  Indeed,
the pair yields in the invariant-mass range between 0.15 and
0.6~GeV/c$^2$, i.e. just below the $\rho$ meson pole mass, still
remain to be explained~\cite{Cassing:1999es,ernst,fuchs,cozma}.
Recently, the NA60 collaboration has published new data on dimuon
production which allow a determination of the in-medium spectral
function of the $\rho$ meson in 158~AGeV In+In collisions
\cite{Dam2005ni}.

The High-Acceptance DiElectron Spectrometer HADES at GSI, Darmstadt,
operates in the SIS/Bevalac energy regime of 1-2 AGeV. In this
Letter we report on the first measurement of inclusive electron-pair
production in $^{12}$C+$^{12}$C collisions at a kinetic beam energy
of 2~AGeV.

\begin{figure}
  \vspace*{0.3cm}
  \mbox{\epsfig{figure={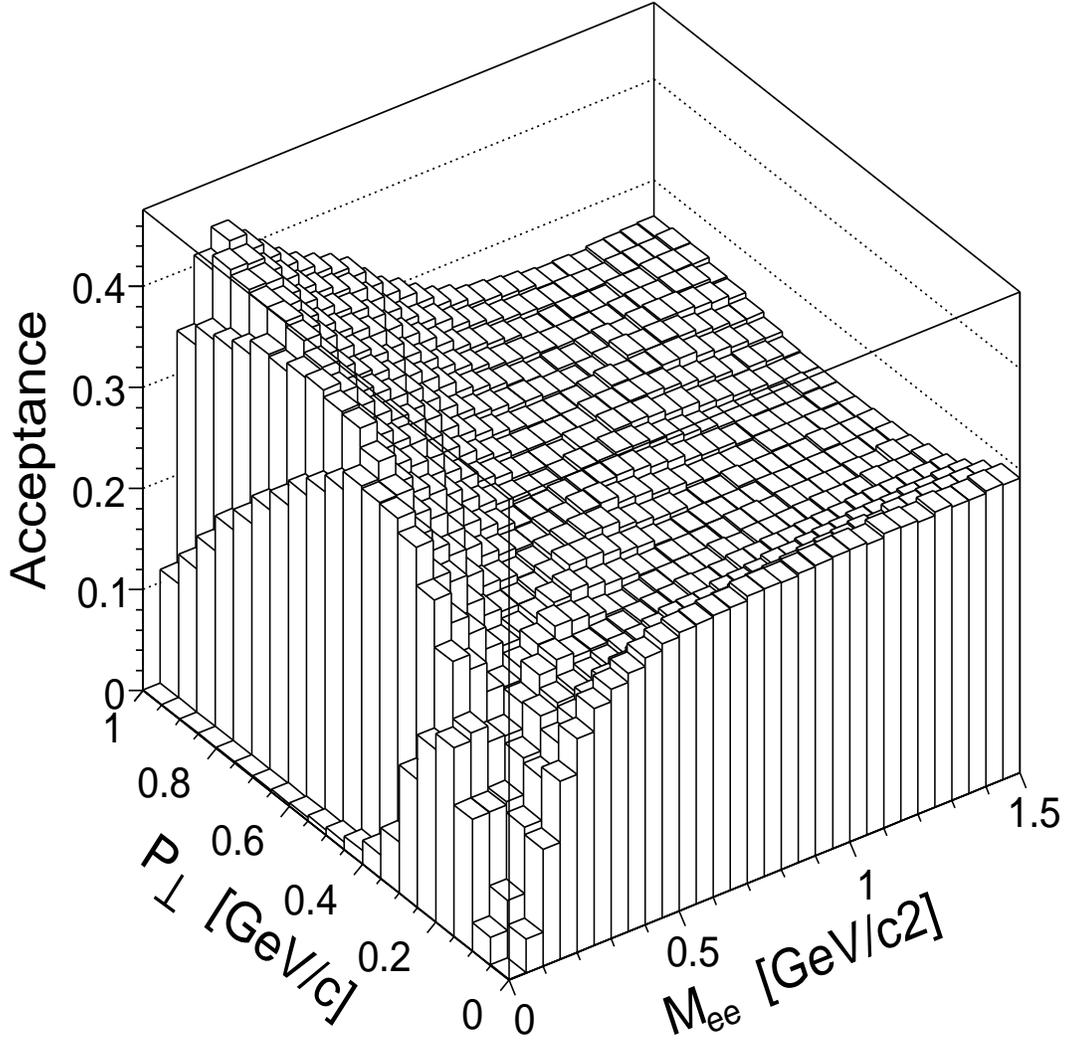},bbllx=0,bblly=0,bburx=540,bbury=490,width=0.9\linewidth,height=0.9\linewidth}}
  \caption{\small Geometrical acceptance of the HADES spectrometer for $e^+e^-$ pairs with
  laboratory opening angle $\theta_{e+e-}>9^0$ as a function of their invariant mass and
  transverse momentum.
  \label{hadspek} }
  \vspace*{-0.1cm}
\end{figure}

A carbon beam of $10^6$ particles/s was incident on a 2-fold
segmented carbon target with a thickness corresponding to
$2\times2.5\%$ interaction lengths. The HADES spectrometer,
described in detail in Refs. \cite{NIM,hades}, consists of a 6-coil
toroidal magnet centered on the beam axis and six identical
detection sections located between the coils and covering polar
angles between $18^\circ$ and $85^\circ$.  In the measurement
presented here, each sector was composed of a gaseous Ring-Imaging
Cherenkov (RICH) detector, two planes of Mini-Drift Chambers (MDC-I
and MDC-II) for track reconstruction and a Time-Of-Flight wall
(TOF/TOFino) supplemented at forward polar angles with Pre-SHOWER
chambers.  The interaction time was obtained from a fast diamond
start detector located upstream of the target. This geometry results
in a smooth dielectron acceptance, shown in ~Fig. \ref{hadspek} as a
function of invariant mass and transverse momentum, averaged over
the rapidity range $0<y<2$.  The data readout was started by a
first-level trigger (LVL1) decision, requiring a charged-particle
multiplicity $MUL \geq 4$ in the TOF/TOFINO detectors, accepting
$60\%$ of the total cross section. It was followed by a second-level
trigger (LVL2) requesting at least one electron track.  With this
trigger condition, a 10-fold pair enrichment at a pair efficiency of
$\geq 92\%$ was achieved. Furthermore, the LVL2 introduced no bias
on the shapes of measured pair distributions, as checked by a direct
comparison to the unbiased LVL1 events.

The results presented below were obtained from events with a
positive LVL2 decision with a total statistics corresponding to
$6.5\times10^8$ LVL1 events. The electron track reconstruction
proceeded in four steps (see also \cite{hades}):

(1) The RICH ensures electron detection for momenta $p>$ 0.05 GeV/c.
From the measured ring positions, polar ($\theta$) and azimuthal
($\phi$) angles are calculated assuming electron emission from the
target.

(2) The $\theta$, $\phi$ values of electron candidates in the RICH
were correlated with the track angles reconstructed in the MDC
within windows corresponding to $\pm$2~standard deviations, as
derived from the electron distributions.

(3) Candidate tracks were then matched with those hits in the TOF or
TOFINO/Pre-SHOWER detectors fulfilling electron conditions, i.e.\
(a) a particle velocity of $\beta=1\pm3\sigma_{\beta}$, with
$\sigma_{\beta}$ given by the time-of-flight resolution, and (b) an
electromagnetic shower signal in the Pre-SHOWER.

(4) Finally, the track momentum was determined by a fit of an
appropriate track model to the reconstructed hit positions making
use of the deflection in the known magnetic field.

The identified single-electron tracks were combined into
opposite-sign pairs from which invariant-mass distributions, with
the mass resolution $\sigma_{M_{ee}}/M_{ee}=9\%$ at $M_{ee}=0.8$
GeV/c$^2$, were built. Many of these $e^+e^-$ pairs, however,
represent combinatorial background (CB) which has to be reduced and
subtracted.  The CB is mostly due to uncorrelated electrons from
$\pi^0\to\gamma\gamma$ decays followed by photon conversion, either
in the target or in the RICH radiator, and/or from $\pi^0\to
e^+e^-\gamma$ Dalitz decays. Such pairs have small opening angles
and often produce partially overlapping tracks in the MDC.  They are
rejected efficiently by applying conditions on the opening angle,
$\theta_{e+e-}>9^{\circ}$, and on the fit quality ($\chi^2$) of the
reconstructed track segments, removing $95\%$ of the conversion
pairs while reducing the dielectron signal with $M_{ee}>0.15$
GeV/c$^2$ by less than $10\%$.

\begin{figure}[htb]
  \mbox{\epsfig{figure={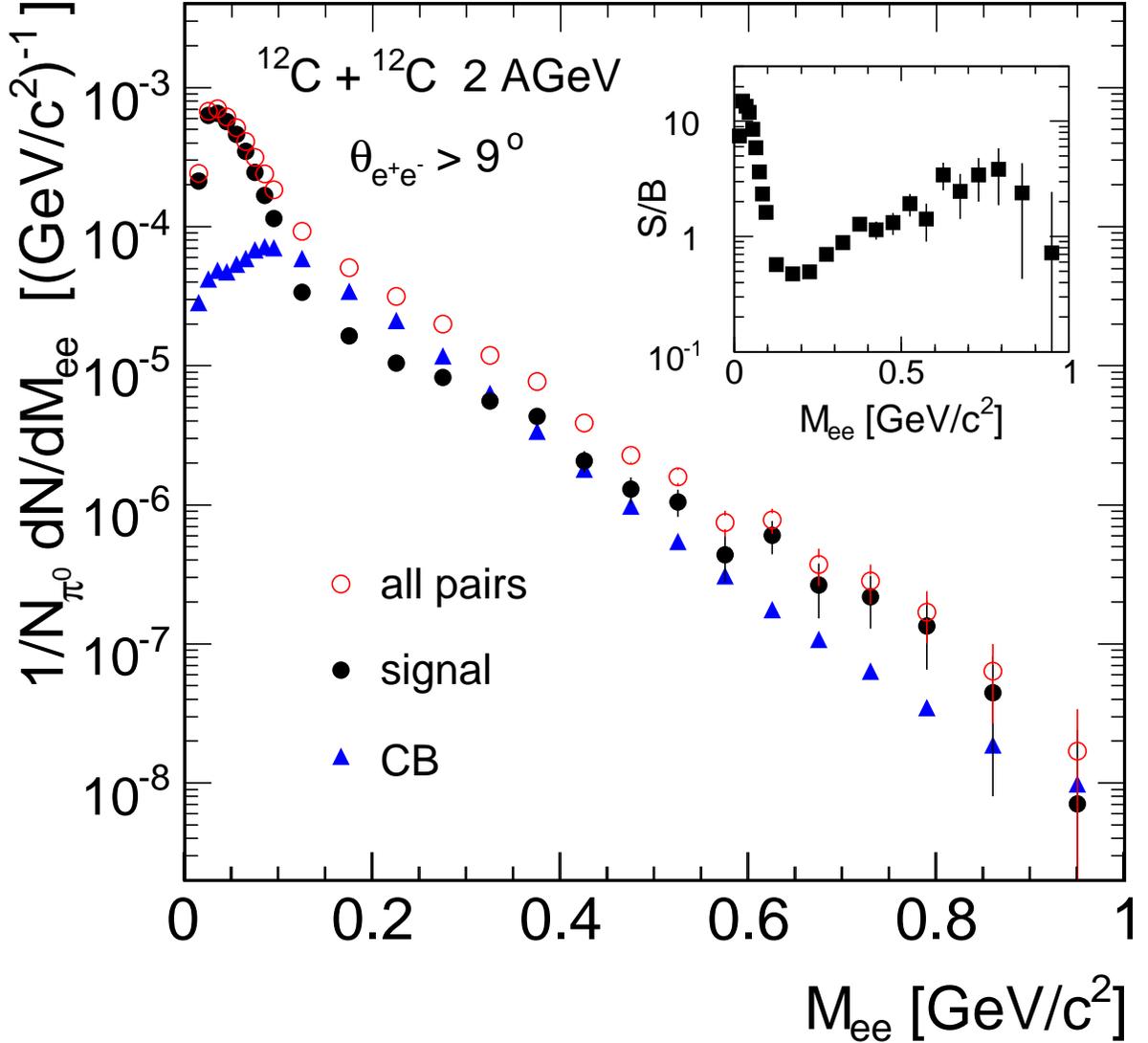},bbllx=0,bblly=0,bburx=520,bbury=548,
                width=0.9\linewidth,height=0.95\linewidth}}
  \vspace*{-0.5cm}
  \caption{\small Reconstructed (not efficiency corrected) $e^+e^-$ invariant-mass
  distributions (open circles) for pairs with opening angles $\theta_{e+e-}>9^o$.
  The signal distribution (dots) was obtained by subtracting the
  combinatorial background (full triangles). Errors indicated are statistical only.
  Inset: Signal-to-background (S/B) ratio versus invariant mass.
  \label{invmas}}
  \vspace*{-0.1cm}
\end{figure}

Fig.~\ref{invmas} shows the resulting $e^+e^-$ invariant-mass
distribution, its signal part, the CB and the resulting signal/CB
ratio.  Like-sign $e^+e^+$ and $e^-e^-$ pairs were formed and
subjected to the same selection criteria as the opposite-sign pairs.
From the reconstructed like-sign invariant-mass distributions,
$dN^{++}/dM_{ee}$ and $dN^{--}/dM_{ee}$, the respective CB
distribution was calculated as $N_{CB} = 2\sqrt{N^{++}N^{--}}$.  For
masses $M_{ee}>0.5$ GeV/$c^2$, where statistics is smaller, the CB
was obtained by an event mixing procedure.  Uncorrelated
opposite-sign $e^+e^-$ pairs were formed from different events but
originating from reactions in the same target segment.  It was
verified that for $M_{ee}>0.15$ GeV/c$^2$ the CB distributions
obtained from the event mixing and the like-sign pairs agree within
$10\%$. All distributions were normalized to the number of neutral
pions $N_{\pi^0}$ (see below).  In total $\simeq23000$ signal pairs
($\simeq2000$ with $M_{ee}>0.15$ GeV/c$^2$) were reconstructed.

We corrected the spectra for detector and reconstruction
inefficiencies by Monte-Carlo simulations embedding electron tracks
with uniform $1/p$ and isotropic angular distributions into
$^{12}$C+$^{12}$C events generated with the UrQMD transport model
\cite{bleicher}. Resulting events were digitized and processed
through the same analysis chain as the measured data. The
single-electron efficiencies, $\epsilon_{\pm}$, were calculated as a
function of charge ($\pm$), momentum ($p$), polar ($\theta$) and
azimuthal ($\phi$) emission angles. The data were corrected on a
pair-by-pair basis with the weighting factor $1/E_{+-}$, with
$E_{+-}=\epsilon_+ \cdot \epsilon_-$ for given electron momenta and
emission angles. The CB was treated likewise and subtracted, as
described above, to obtain the efficiency-corrected pair signal
distribution.  This prescription relies on the assumption that the
single-leg efficiencies are independent, as was carefully checked in
our simulations and proven to be valid within $15\%$ for pairs with
opening angles $\theta_{e+e-}>9^0$.  The geometrical pair acceptance
of the HADES detector was obtained in analogy to the pair efficiency
as the product of two single-electron acceptances
$A_{\pm}(p,\theta,\phi)$. The resulting matrices, together with a
momentum resolution function, constitute the HADES acceptance filter
(available upon request).  We made no attempt to extrapolate the
measured dielectron yields to the full solid angle.

\begin{figure}[htb]

  \mbox{\epsfig{figure={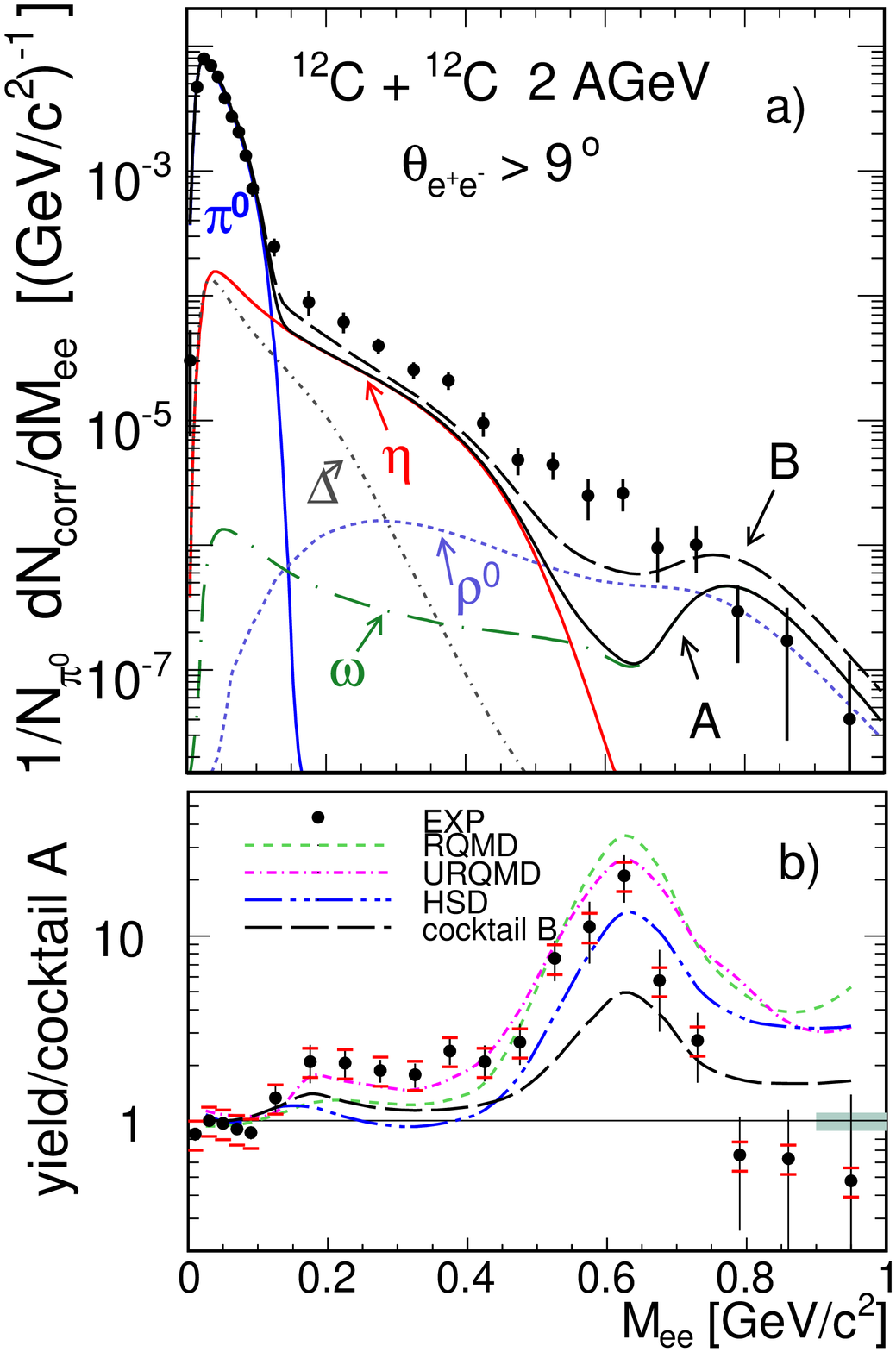}, bbllx=1,bblly=1,bburx=540,bbury=760,
  width=0.9\linewidth,height=1.1\linewidth}}
  \vspace*{-0.1cm}

  \caption[]{
  (a) Efficiency- and background-corrected $e^+e^-$ invariant-mass
  distribution for $\theta_{e+e-}>9^o$ (symbols) compared to a thermal
  dielectron cocktail of free $\pi^0$, $\eta$ and $\omega$ decays
  (cocktail A, solid line), as well as including $\rho$ and $\Delta$ resonance decays
  (cocktail B, long-dashed line).  Only statistical errors are shown.
  (b) Ratio of data and cocktail A (dots), compared to ratios
  of various model calculations and cocktail A.  All calculations
  have been filtered and folded with the HADES acceptance and mass
  resolution.  Statistical and systematic errors of the measurement
  are shown as vertical and horizontal lines, respectively.
  The overall normalization error of 11\% is depicted by the shaded area.
  }

  \label{signal}

\end{figure}

Fig.~\ref{signal}a shows the $e^+e^-$ invariant-mass distribution of
the signal pairs after efficiency correction and normalized to the
average number of charged pions
$N_{\pi}=\frac{1}{2}(N_{\pi^+}+N_{\pi^-})$.  The latter were
identified in HADES by means of the time-of-flight measurement
\cite{hades} and their yield was extrapolated to full solid angle,
taking into account our measured angular distributions found to be in
agreement with UrQMD calculations.  In the isospin-symmetric system
$^{12}$C+$^{12}$C, $N_{\pi}$ is in fact a good measure of the
$\pi^0$ yield.  This way of normalizing the pair spectra compensates
to first order the bias caused by the implicit centrality selection
of our trigger.  Indeed, simulations based on UrQMD events indicate
that LVL2 events have an average number of participating nucleons
$A_{part}=9.0$, instead of 6 for true minimum-bias events.  The pion
multiplicity per number of participating nucleons
$M_{\pi}/A_{part}=0.137\pm0.015$ obtained in our experiment agrees
with previous measurements of charged and neutral pions
\cite{taps1,kaos} within the quoted error of $11\%$. The error is
dominated by systematic uncertainties in the acceptance and
efficiency corrections of the charged-pion analysis and represents
our overall normalization error.

We first compare our results with a pair cocktail (cocktail A)
calculated from free $\pi^0$, $\eta$ and $\omega$ meson decays
only (Fig~\ref{signal}a).  This cocktail aims at representing all
contributions emitted after the chemical freeze-out of the
fireball.  While the first two sources are directly constrained by
data \cite{taps1}, the production rate of the $\omega$ meson is
taken from an $m_{\perp}$-scaling ansatz \cite{bratkovskaya}.
In our event generator (PLUTO \cite{PLUTO}) meson production was
hence modeled assuming emission from a thermal source with a
temperature $T=80$~MeV, but no radial expansion velocity ($\beta_r$=0).
Furthermore, for the $\pi^0$ mesons, an anisotropic angular
distribution of the type
$dN/d\cos(\theta_{CM}) \sim 1+a_2 \cdot cos^2(\theta_{CM})$ with
$a_2=0.7$ was used, as deduced from our charged pion analysis. The
accepted lepton-pair yield was found to change by less than 14\%
when varying the source parameters over a broad range: $\beta_r$
(0-0.3), $a_2$ (0-1.0) and $T$ (50-90 MeV).

While experimental data and simulated cocktail A (solid line in
Fig.~\ref{signal}a) are in good agreement in the $\pi^0$ region, the
coctail undershoots the data for $M>0.15$ GeV/c$^2$ and clearly
calls for additional sources. Such contributions are indeed expected
from the decay of short-lived resonances, mainly the $\Delta(1232)$
and the $\rho$, excited in the early phase of the collision.  To
include in our cocktail pairs from $\Delta^{0,+}\to Ne^+e^-$ decays,
we assumed that the $\Delta$ yield scales with the $\pi^0$ yield and
employed a calculated decay rate \cite{ernst}.  To add the $\rho$
meson we used a similar prescription as for the $\omega$. For this
broad resonance ($\Gamma_{\rho}=150$ MeV), $m_{\perp}$ scaling
strongly enhances the low-mass tail, resulting in the skewed shape
visible in Fig.~\ref{signal}a.  The long-dashed line in this figure
shows the comparison of the full thermal cocktail (cocktail B) with
our data. As expected, the simulated yield above 0.15 GeV/c$^2$ is
now increased and, in particular, the high-mass region is populated
with dielectrons from $\rho \to e^+e^-$ decays, but the calculation
still falls short of reproducing the data.

To discuss in more detail the excess pair yield, we show in
Fig.~\ref{signal}b the ratio of the data and cocktail A. Statistical
and systematical errors originating from the CB subtraction,
efficiency corrections and normalization, are shown separately. In
the intermediate mass range of $0.15 - 0.50$ GeV/c$^2$, the
enhancement factor above the dominant $\eta$ contribution amounts to
$F(2.0) = Y_{tot}(2.0)/Y_{\eta}(2.0)=2.07 \pm 0.21(stat) \pm 0.38
(sys)$. Assuming that the excess pairs have in this mass region an
overall acceptance close to the one of eta Dalitz pairs, one can
compare $F(2.0)$ to the enhancement factor measured in C+C by DLS at
a beam energy of 1.04~AGeV \cite{dls}. Using the DLS data and an
appropriately filtered PLUTO cocktail generated for this bombarding
energy, we obtain a factor of $F(1.04)=6.5 \pm 0.5(stat) \pm
2.1(sys)$. Knowing that, going from 1.04 to 2 AGeV, inclusive $\eta$
production in C+C collisions increases by a factor
$Y_{\eta}(2.0)/Y_{\eta}(1.04)=13\pm 3$ \cite{taps1, taps2}, the
energy scaling factor of the excess pair yield
$Y_{exc}(2.0)/Y_{exc}(1.04)$, where $Y_{exc}=Y_{tot}-Y_{\eta}$, can
be deduced from the two enhancement factors $F(2.0)$ and $F(1.04)$.
It follows that $Y_{exc}(2.0)/Y_{exc}(1.04)=2.5\pm 0.5 (stat) \pm
1.5 (sys)$. This energy scaling is remarkably similar to the known
scaling of pion production, i.e.
$Y_{\pi}(2.0)/Y_{\pi}(1.04)=2.3\pm0.3$ \cite{taps1, taps2}.  It
suggests that the pair excess is indeed driven by pion dynamics,
involving e.g. $\Delta$ and $\rho$ excitations.

At still higher masses the ratio of data and cocktail A develops a
pronounced maximum around $M_{ee}\sim 0.6$ GeV/c$^2$, mainly due to
the lack of $\rho$ decays in cocktail A.  Especially the mass region
between the $\eta$ and the $\omega$ pole is expected to be dominated
by the thermally populated low-mass tail of the broad $\rho$
resonance. Cocktail B (long-dashed line in Fig. \ref{signal}b),
which includes $\rho$ and $\Delta$ decays, shows indeed an
enhancement, but still does not fully explain the observed pair
yield.

Beyond the cocktail calculations discussed above we have also made a
comparison with various transport models.  The latter, besides
offering a realistic treatment of collisions dynamics, also handle
the propagation of broad resonances, related off-shell effects and
multi-step processes, all known to play a crucial role at our
bombarding energy.  Hence, dielectron distributions were calculated
with the HSD \cite{Cassing:1999es}, RQMD \cite{cozma} and UrQMD
\cite{bleicher} transport models (assuming vacuum spectral
functions).  All calculations were filtered with the HADES
acceptance and normalized to their respective $\pi^0$ yields. The
respective ratios of calculated dielectron yields to our cocktail A
are shown as curves in Fig.~\ref{signal}b.  Due to the
normalization, all models agree in the $\pi^0$ mass region and give
ratios consistent with unity.  In the intermediate mass region all
transport models fall short, with UrQMD coming closest to the data.
As checked, all models agree on the $\eta$ contribution itself
within $20\%$, and the discrepancy can be traced to differences in
their respective treatment of population and decay of the baryonic
resonances \cite{Cassing:1999es,cozma,bleicher}.  At higher masses,
$M_{ee}>0.5$ GeV/c$^2$, all models qualitatively reproduce the trend
given by the data, but differ in the yield due to different
amplitudes used for the couplings to intermediate resonances
\cite{cozma,bleicher}. All transport models consistently
overestimate the pair yield at the $\rho,\omega$ meson poles.  Final
conclusions can be drawn only when more refined calculations with
in-medium spectral functions become available.

In summary, we report on the first measurement of inclusive
dielectron production in $^{12}$C+$^{12}$C collisions at
$E^{kin}_{beam}$=2~AGeV, spanning five orders of magnitude in yield.
At low masses, i.e.\ $M_{ee}<0.15$ GeV/c$^2$, the pair yield is in
agreement with the known $\pi^0$ production and decay probabilities.
For $0.15$ GeV/c$^2$ $<M_{ee}<0.5$ GeV/c$^2$ it exceeds expectations
based on the known production and decay rates of the $\eta$ meson by
$2.07 \pm 0.21(stat)\pm 0.38 (sys)$.  This pair yield excess is
consistent with that measured by DLS at 1.04 GeV if its energy
scaling is similar to that of pion production.  Additional sources
associated with the radiation from the early collision phase
($\Delta^{0(+)}\to Ne^+e^-$, $\rho \to e^+e^-$) are needed to
account for the excess observed for $M>0.15$ GeV/$c^2$. However,
transport calculations based on vacuum spectral functions only still
fail to quantitatively describe the excess yield in the full
invariant-mass range.

The collaboration gratefully acknowledges the support by BMBF grants
06TM970I, 06GI146I, 06F-140, and 06DR120 (Germany), by GSI
(TM-FR1,GI/ME3,OF/STR), by grants GA CR 202/00/1668 and GA AS CR
IAA1048304 (Czech Republic), by grant KBN 1P03B 056 29 (Poland), by
INFN (Italy), by CNRS/IN2P3 (France), by grants MCYT
FPA2000-2041-C02-02 and XUGA PGID T02PXIC20605PN (Spain), by grant
UCY-10.3.11.12 (Cyprus), by INTAS grant 03-51-3208 and by EU
contract RII3-CT-2004-506078.


\end{document}